\documentclass[%
 preprint,
 amsmath,amssymb,
 aps,
]{revtex4-2}

\usepackage{mathtools}
\usepackage{graphicx}
\usepackage{bm}
\usepackage[mathlines]{lineno}
\linenumbers\relax 

\begin{document}
\nolinenumbers

\preprint{Preprint}

\title{Three-dimensional internal flow evolution of an evaporating droplet and its role in particle deposition pattern}

\author{Jiaqi Li}
\author{Jiarong Hong}%
 \email{jhong@umn.edu}
\affiliation{Saint Anthony Falls Laboratory, University of Minnesota, Minneapolis, Minnesota, 55455, USA}
\affiliation{Department of Mechanical Engineering, University of Minnesota, Minneapolis, Minnesota, 55455, USA}

\date{\today}

\begin{abstract}
The internal flow within an evaporating sessile droplet is one of the driving mechanisms that lead to the variety of particle deposition patterns seen in applications such as inkjet printing, surface patterning, and blood stain analysis. Despite decades of research, the causal link between droplet internal flow and particle deposition patterns has not been fully established. In this study, we employ a 3D imaging technique based on digital inline holography to quantitatively assess the evolution of internal flow fields and particle migration in three distinct types of wetting droplets, namely water droplets, sucrose aqueous solution droplets, and SDS aqueous solution droplets, throughout their entire evaporation process. Our imaging reveals the three-stage evolution of the 3D internal flow regimes driven by changes in the relative importance of capillary flow, Marangoni flow, and droplet boundary movement during evaporation. Each droplet type exhibits unique dynamics: water droplets experience competition between capillary and Marangoni flows; sucrose solution droplets are dominated by capillary flow; while SDS solution droplets initially show a strong Marangoni flow that gradually diminishes. The migration of individual particles from their initial locations to deposition can be divided into five categories, with some particles depositing at the contact line and others inside the droplet. In particular, we observe the changing migration directions of particles due to competing Marangoni and capillary flows during droplet evaporation. We further develop an analytical model that predicts the droplet internal flow and deposition patterns and determines the dependence of the deposition mechanisms of particles on their initial locations and the evolving internal flow field. The model, validated using different types of droplets from our experiment and the literature, can be further expanded to other Newtonian and non-Newtonian droplets, which can potentially serve as a real-time assessment tool for particle deposition in various applications.
\end{abstract}

\maketitle


\section{\label{sec:1}Introduction}

Colloidal droplets can leave a large variety of deposition patterns on the substrate after drying like coffee rings and stains left by rain droplets on windows, as commonly observed in daily life. Understanding the underlying mechanisms behind such pattern formation is also of great significance for many practical applications, such as inkjet printing \cite{lim2009,hu2020sa}, surface patterning \cite{van2006}, 3D printing \cite{kong2014}, medical diagnostics of blood diseases \cite{brutin2011}, and bloodstain pattern analysis in forensic science \cite{attinger2013}. Despite many factors like surface adhesion, particle shape, and capillary attraction that can affect deposition patterns \cite{larson2014}, the internal fluid motion of an evaporating droplet (referred to as internal flow hereafter) has been considered as a major driving mechanism that dictates the migration of particles during evaporation and leads to different deposition patterns \cite{deegan1997,deegan2000,fischer2002}.

Consequently, a number of studies have been conducted to investigate the internal flow of evaporating droplets and corresponding deposition patterns under various conditions using theoretical approaches, numerical simulations, and laboratory experiments, as summarized in several review papers \cite{larson2014, brutin2015droplet, parsa2018, gelderblom2022evaporation, wilson2023}. Specifically, the early study by Deegan et al. \cite{deegan1997} postulated that the higher evaporative flux near the contact line of a droplet can induce an outward radial flow (referred to as capillary flow) and migration of particles toward the contact line, leading to the formation of “coffee-ring” deposition. Subsequently, such a capillary flow was investigated using analytical methods and numerical simulation by Hu and Larson \cite{hu2005a}. The same researchers also incorporated the Marangoni flow, i.e., the surface tension-driven, recirculating flow due to thermally or chemical gradient across a droplet surface \cite{ehrhard1991,fanton1998}, into the analytical and numerical modeling of internal flows and quantified the resulting recirculatory flow motion inside an evaporating droplet \cite{hu2005b}. Later, they conducted simulations of particle motion driven by such internal flows and predicted a distinct center deposition pattern for an octane droplet influenced by a strong thermal Marangoni effect, which was validated by their experiments \cite{hu2006}. In addition, using an analytical approach, Masoud and Felske \cite{masoud2009} showed that the contact line movement during droplet evaporation can alter internal flow by reversing the outward flow for uniform flux across the droplet–air interface in comparison to those present in the earlier studies using fixed boundary conditions. Such effect was also considered in the numerical simulation \cite{bhardwaj2009}. Their study demonstrated the initial development of Marangoni flow (in water droplet) and later dominant capillary flow as the boundary condition changes and predicted no significant changes in the deposition patterns for the simulated water ("coffee-ring" deposition) and isopropanol (volatile; center deposition) droplets. Using spectral radar optical coherence tomography (SR-OCT) measurements, Manukyan et al. \cite{manukyan2013} captured snapshots of 3D particle migration inside droplets of paint-water mixtures on hydrophilic and hydrophobic surfaces and suggested the reverse of 3D internal flow circulation causes the formation of “coffee-ring” and center deposition in these two types of surfaces, respectively. Larson \cite{larson2014} reviewed the various internal flow regimes and deposition patterns under different conditions and pointed out the lack of experimental data to support mechanisms proposed to link internal flow and particle depositions. This limitation is then highlighted by Kim et al. \cite{kim2016} using the planar particle image velocimetry (PIV), where they observed the fast transition of internal flow regimes within the first 10 seconds of droplet evaporation and found that slight differences in the chemical compositions (e.g., surfactant concentration) of the whiskey droplets would lead to drastically different internal flow evolution and deposition patterns. More recently, Marin, Rossi, and their colleagues \cite{marin2016surfactant,marin2019solutal,rossi2019interfacial} have advanced droplet internal flow measurements by applying the 3D astigmatism particle tracking velocimetry (APTV) method. Their measurements quantified the 3D droplet internal flow dynamics within droplets with different additives, especially near the droplet air interface \cite{marin2016surfactant,rossi2019interfacial}, and provided a potential explanation for the "coffee-ring" deposition due to interfacial flow \cite{marin2019solutal}. However, these experimental works have not provided detailed particle migration motions throughout the evaporation process through their particle tracking. Such information is critical to elucidate the direct linkage between internal flow evolution and the change of deposition patterns. 

In this study, we tackle the technical difficulties and elucidate the physical mechanisms of the particle deposition by investigating the evolving internal flow and particle transport inside evaporating droplets using the digital inline holography (DIH). DIH has emerged as a unique technique at the beginning of the 21st century for 3D imaging of various fluid flows and particle transport \cite{xu2001,katz2010}, including flow and particle transport in microfluidic systems \cite{seo2014,qin2019}, turbulent channel flow over smooth and rough surfaces \cite{sheng2009,toloui2017}, and the locomotion of microorganisms \cite{sheng2007,su2012,molaei2014}. Compared to the astigmatic particle tracking velocimetry (APTV), DIH has a larger depth of field (DOF), and it exhibits superior depth sensitivity when determining the longitudinal location of particles. Additionally, DIH is able to accommodate a higher particle concentration, ensuring enhanced resolution for flow field visualization. With the DIH measurements, the 3D locations of the suspended particles can be extracted and tracked from the beginning of the evaporation process to their final deposition. By integrating particle tracking velocimetry (PTV) with DIH, our measurements shed light on the intricate physical processes governing internal flow dynamics and particle deposition mechanisms during droplet evaporation. In Section \ref{sec:2}, we detail our experimental setup, material preparation, and data processing techniques. Section \ref{sec:3} presents our findings, where we comprehensively analyze the evaporation processes of three distinct droplet systems. We explore the evolution of internal flow and particle migration, and examine the resulting deposition patterns. The underlying connections between these phenomena are then established and discussed, offering insights and hypotheses on the causal link between the internal flow dynamics of droplets and their final particle deposition patterns. To corroborate our experimental findings, we further introduce a predictive model. Finally, conclusions and discussion follow in Section \ref{sec:4}.

\section{\label{sec:2}Materials and methods}
\subsection{\label{sec:21}Material preparation and experiment}

\begin{figure}
\includegraphics{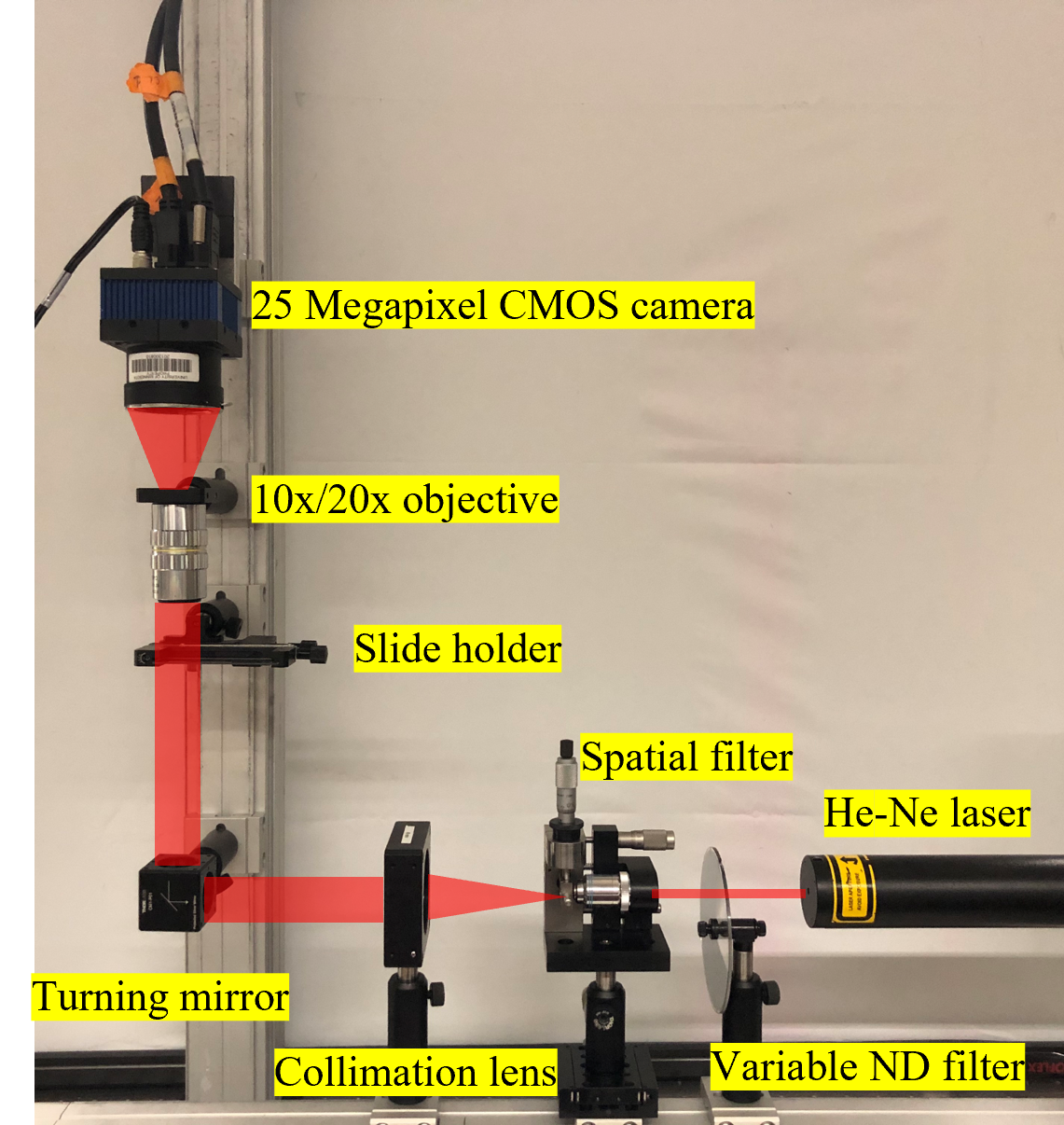}
\caption{The experimental setup of the digital inline holography (DIH) system.}
\label{fig:1}
\end{figure}

To study the internal flow and particle motions during the evaporation of droplets on the hydrophilic substrate, we conduct experiments using digital inline holography (DIH) and particle tracking velocimetry (PTV). The experiment setup (Fig. \ref{fig:1}) consists of a He-Ne laser (633 nm wavelength) as the illumination source, followed by a spatial filter (an objective lens for focusing and expanding the beam and a pinhole for filtering). The expanding beam is then collimated by a convex lens with 35 mm in diameter and turned to the vertical direction for bottom view imaging by a 90\textdegree{} turning mirror in a 30 mm cage with 25 mm circular opening. A slide holder is placed between the turning mirror and a magnifying objective lens within the sampling volume. We use a 25-megapixel (5000 $\times$ 5000) complementary metal–oxide–semiconductor (CMOS) camera (Vieworks Co. Ltd) to record the holograms (i.e., the fringe patterns generated by the interference between the scattered signals from the tracer particles and the unscattered portion of the illumination light source) of the evaporating droplets at 20 frames per second. The holograms contain information on the 3D position (longitudinal and lateral) of the particles in the sample volume. Such data can be extracted through holographic reconstruction, and particle tracking velocimetry (PTV) can then be applied to track the 3D motions of the reconstructed particles. Two objective lenses with different magnifications (10$\times$ and 20$\times$) are used to capture the internal flow of the whole droplet under 0.52 µm/pixel spatial resolution and detailed particle movements that lead to deposition throughout the evaporation process under 0.3 µm/pixel resolution, respectively. We utilize side-view imaging using a Sony A7RII camera to measure the temporal variations of the contact angle in addition to the DIH imaging.

As reported in previous literature \cite{morales2013,kim2016,shimobayashi2018}, SDS (sodium dodecylsulfate) and sucrose modify the surface tension, thus leading to different deposition patterns compared to that of the water droplets. Specifically, as documented by \cite{mysels1986surface}, the surface tension of the aqueous SDS solution decreases with its concentration, up to its critical micellar concentration of 8.3 mM. While the surface tension of the aqueous sucrose solution rises as its concentration increases \cite{butler1923ccxxix}. Additionally, the viscosity of sucrose \cite{vand1948viscosity, hidayanto2010measurement} and SDS \cite{kodama1972second, ruiz2018coarse} solutions in water increases with concentration, rising by up to 10$\%$ within the tested range in our experiments. Considering the different effects additives can have on the evaporating droplets, we test multiple droplet systems for the experiment. Distilled water, SDS (Sigma-Aldrich, reagent grade $\geq 98.5 \%$) aqueous solution, and sucrose (Sigma-Aldrich, reagent grade) aqueous solution are used. We prepare three concentrations for both the SDS solution and sucrose solution. For the SDS solution, 35 mmol/L (mM for short, as M represents molarity mol/L), 17 mM, and 1.7 mM concentrations are selected, and for sucrose, 10 mM, 1 mM, and 0.1 mM concentrations are chosen for the experiment. The droplets are deposited on brand-new glass slides (AmScope). Specifically, we used high-quality ethanol (200 Proof, 100\%, Decon Labs, Inc.) to clean these glass slides, and air-dry them using a duster, avoiding mechanical rubbing to prevent potential scratches. This cleaning method maintains the surface smoothness, ensuring consistent hydrophilicity.  The deposited droplets have a volume of 0.5 µL and an initial diameter of 2 mm, seeded with 10 µm polystyrene particles (Kisker Biotech, diluted to 0.005w/w$\%$) to track the fluid motion. These tracer particles have a density close to that of water, rendering them neutrally buoyant within the droplets. The experiments are conducted in an environment with a 21 \textdegree{C} temperature and $45\% \pm 5\%$ relative humidity. The total time of evaporation is around 5 minutes on average, which establishes the time scale for the internal flow. Given this time scale, the tracer particles exhibit a Stokes number on the order of $10^{-9}$, making them ideal tracers with minimum inertial effect.  The recording starts around 20 seconds (a conservative estimate) after the droplet is placed on the glass slide until the end of the evaporation process. This initial waiting time is required to carefully deposit the droplet and subsequently locate it within the imaging field of view, ensuring that the droplet contact line is as round as possible. Moreover, this early stage of internal flow development has a minimal impact on the final deposition pattern. In addition, a soft plastic shield is securely taped to the camera casing, encircling both the objective lens and the glass slide. This setup is specifically designed to ensure that the droplet evaporation remains unaffected by any potential background flows present in the laboratory environment.

\subsection{\label{sec:22}Experiment data processing}

\begin{figure}
\includegraphics{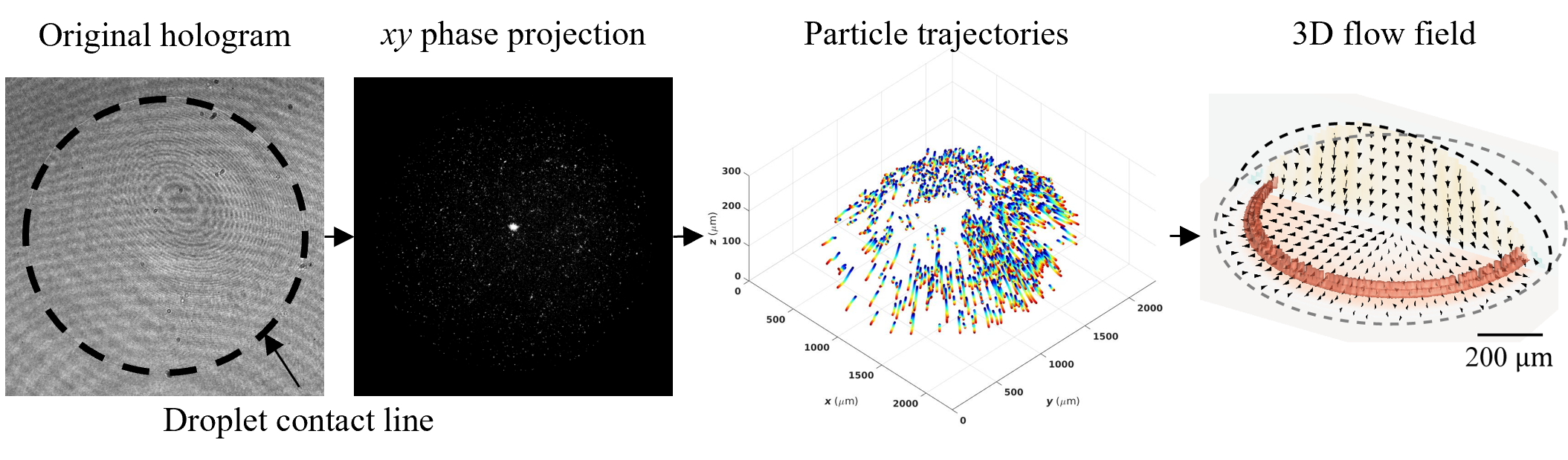}
\caption{Illustrations of the experimental data processing procedures, including hologram reconstruction, particle tracking velocimetry, converting Lagrangian velocities to Eulerian flow field.}
\label{fig:2}
\end{figure}

The data processing procedures are illustrated in Fig. \ref{fig:2}. The holograms are first enhanced by subtracting their corresponding Gaussian blurred images considering the changing boundary condition and slow fluid motion. The 3D locations of the tracer particles are obtained from the reconstructed 3D optical field from the enhanced holograms using the regularized inverse holographic volume reconstruction (RIHVR) method by \cite{mallery2019}. The RIHVR approach for the volumetric reconstructions of particle fields can be described by the equation:
\begin{equation}
\hat{x}=\underset{x}{\operatorname{argmin}}\left\{\|H x-b\|_2^2+\lambda R(x) \equiv f(x)+g(x)\right\}
\end{equation}
where $x$ is the optimal particle field to be found that iteratively minimizes the difference between the real hologram b and the estimated hologram by transforming the particle field using an operation $H$. In the transformation $H$, the Rayleigh-Sommerfeld diffraction kernel is used. The $\lambda R(x)$ term is the combined regularization term, the fused lasso regularization, including a $l^1$ term for controlling the sparsity and a total variation regularization term to enforce the smoothness of the reconstructed particle field. The algorithm is implemented with GPU parallel computing, which significantly reduces the computational time as compare to the other compressive holography implementations \cite{brady2009compressive,verrier2016improvement,jolivet2018regularized}. The reconstructed tracer 3D locations are then used in TrackPy \cite{allan2021} to obtain particle trajectories. TrackPy is an open-source Python package for tracking particles in 2D, 3D, and higher dimensions (other parameters and features of the tracked objects) based on the method proposed by \cite{crocker1996}. Furthermore, the evolving 3D flow field can be obtained by interpolation of the scattered velocities from the particle trajectories at a specific time instance to a uniform 3D grid. A 3D Gaussian kernel is then applied to smooth the 3D flow field.

\section{\label{sec:3}Results}
\subsection{\label{sec:31}Varying evaporation behaviors and deposition patterns}

\begin{figure}
\includegraphics{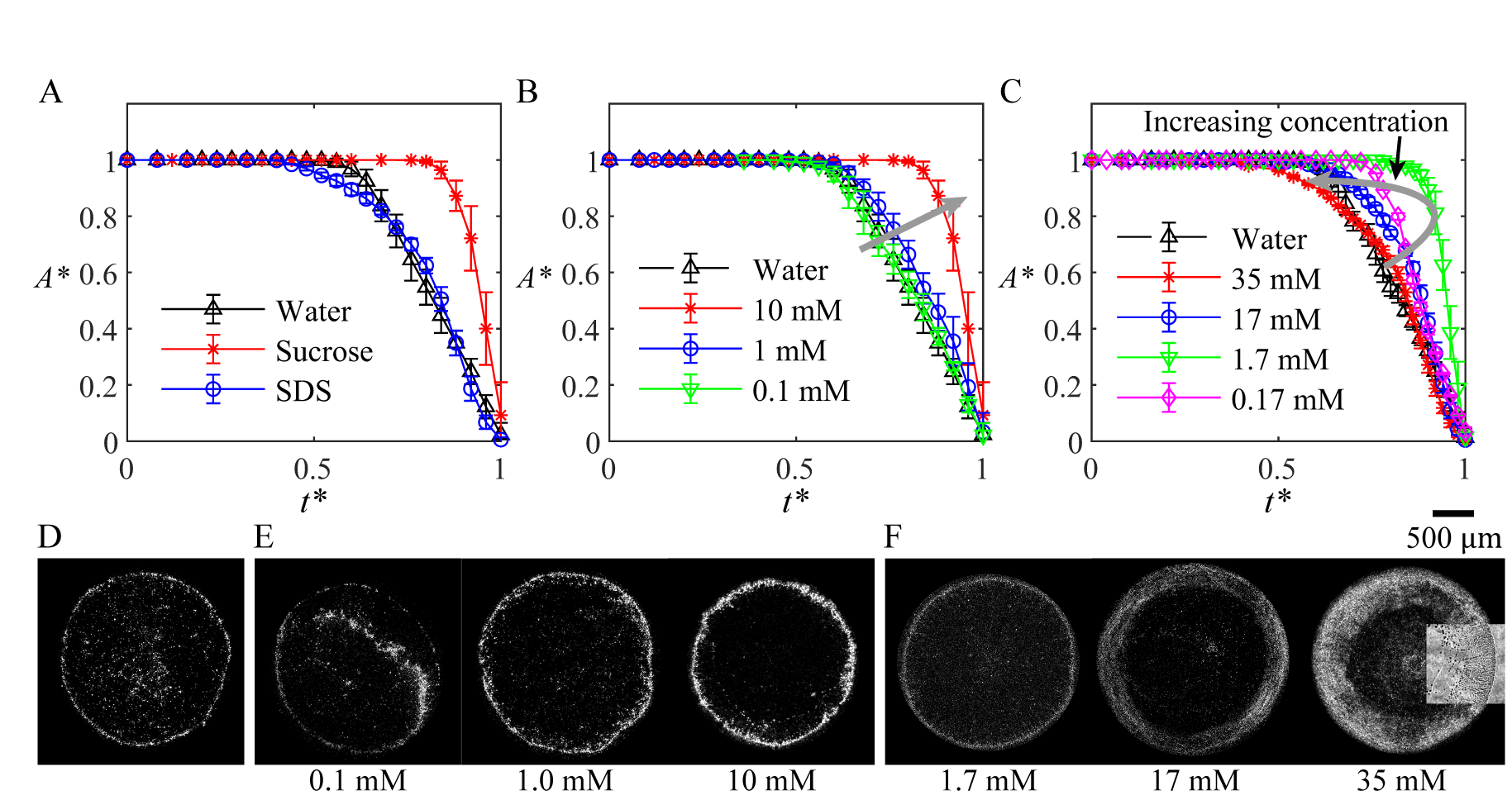}
\caption{Time variation of contact area and the final deposition pattern for droplets with different chemical composition and concentration. (A) Comparison of the time (normalized by the total evaporation time) variation of the droplet contact areas normalized by their initial values (evaporation curve) among the most contrasting cases: evaporating water, 10 mM sucrose solution, and 35 mM SDS solution droplets. (B) The evaporation curves for sucrose solution droplets and (C) SDS solution droplets with different concentrations. (D) Sample binarized phase projections of deposition holograms of a water droplet, (E) sucrose solution droplets, and (F) SDS droplets with different concentrations. (F, \textit{inset}) Original hologram of the crystallization of SDS. The "original hologram" refers to the raw images from which these phase projections are created.}
\label{fig:3}
\end{figure}

The droplets' geometry is determined under the assumption that they conform to a spherical cap shape. The changes of the normalized contact area as functions of time (normalized by the overall drying time) exhibit different features for different droplet systems (Fig. \ref{fig:3} A$\sim$C). These features suggest different evaporation modes that can lead to various deposition patterns illustrated in the reconstructed phase projections from the holograms of multiple tested droplets after drying (Fig. \ref{fig:3} D$\sim$F). These phase projections are derived by first capturing the phase information \cite{latychevskaia2015practical} of the holograms at various reconstruction planes along the $z$ direction (perpendicular to the imaging plane). This collection of phase data is then projected onto a single plane using the maximum phase value, translating the $0 \sim 2 \pi$ scale into a grayscale image. We have neglected the effects of the substrate in the experiments and ensured that the surfaces are hydrophilic by using brand-new and ethanol-washed glass slides for all droplet systems. The resulting initial contact angles are $38^\circ$ for water, $50^\circ$  for sucrose solution, and $22^\circ$  for SDS solution droplets. These angles, reflecting changes in wetting properties due to surfactant and sugar, lead to varying surface areas and, consequently, different overall drying time: 5 min 30 sec for water, 6 min for sucrose solution, and 4 min 30 sec for SDS solution. The discrepancy in surface areas due to these differing contact angles results in up to 17$\%$ variation in overall drying time across the systems, with SDS solution droplets experiencing earlier depinning due to their lower initial angles, and sucrose solution droplets showing a delayed onset of depinning. 

The evaporation process of the water droplet starts with constant contact radius (CCR) mode (the terms corresponding to different evaporation modes were introduced in \cite{picknett1977}), and the flow carries particles to the pinned contact line (referred to as contact line deposition, CLD, hereafter), leading to the “coffee-ring” deposition. It then transitions to a mixed evaporation mode of CCR and constant contact angle (CCA) modes (i.e., slip-stick mode, as shown in Fig. S1 of \textit{SI}) with rapid contact line depinning motion after 60\% of the total evaporation time. Such contact line depinning motion leads to particle deposition inside the initial contact line (referred to as inner deposition, ID, hereafter), leading to a mixture of CLD and ID (Fig. \ref{fig:3}D). Compared to the initial contact line deposition, the inner deposition has a relatively low particle concentration. This low particle concentration results in the deposition at these later contact lines being too sparse to form distinct, visible rings. While for the sucrose solution droplet (Fig. \ref{fig:3}B), a delayed contact line depinning is observed, which results in more prominent “coffee-ring” deposition and less inner deposition than the water droplet (Fig. \ref{fig:3}E). Such a delay is enhanced as the sucrose concentration increases, potentially related to the increased surface tension (i.e., stronger adhesion of the contact line to the substrate) due to higher sucrose concentration, the more pronounced ring deposition, and sucrose crystallization along the contact line.

The evaporation processes and deposition patterns are very different for the SDS solution droplets. As the SDS reduces the droplet surface tension, the contact line depinning starts much earlier than the water and sucrose cases, with the contact line gradually moving inward at an escalating pace. The continuously moving contact line then results in a uniformly scattered deposition with more particles at the center and an outer annulus band caused by crystallization (Fig. \ref{fig:3}F, inset). Moreover, the evaporation behaviors of SDS solution droplets show a non-monotonic trend as the concentration of SDS increases (Fig. \ref{fig:3}C). The increasing concentration from 0 to 1.7 mM delays the contact line depinning, but a further increase in SDS concentration from 1.7 mM to 35 mM leads to earlier depinning with continuous inward contact line motion. However, the deposition patterns of SDS solution droplets still transition from “coffee-ring” deposition of water and low-concentration SDS droplets to more uniform deposition for higher concentration droplets (Fig. \ref{fig:3}F). The solute deposition (crystallization) band grows wider as concentration increases. This non-monotonic dependence on concentration is potentially due to two competing effects influencing the contact line depinning. On the one hand, the particle and solute depositions near the contact line lead to contact line pinning \cite{weon2013}. On the other hand, surface tension decreases with higher concentration up to the critical micellar concentration, leading to smaller initial contact angle, which promotes contact line depinning. As the SDS concentration further increases up to 35 mM, despite almost constant surface tension, the initial contact angle decreases. These droplets' contact angle reaches the critical value ($14^\circ$ based on \cite{hu2005b}) for depinning earlier. For higher SDS concentrations, the influence induced by surface tension and initial contact angle becomes more and more dominant.

\subsection{\label{sec:32}Evolution of internal flow field}

Although the evaporation curves can partly explain the differences in the final particle deposition pattern, the detailed particle motions inside the droplet due to fluid flows remain unknown. In this section, we examine how internal flows of different droplet systems evolve and how they are related to the final deposition patterns.

\begin{figure}
\centering
\includegraphics{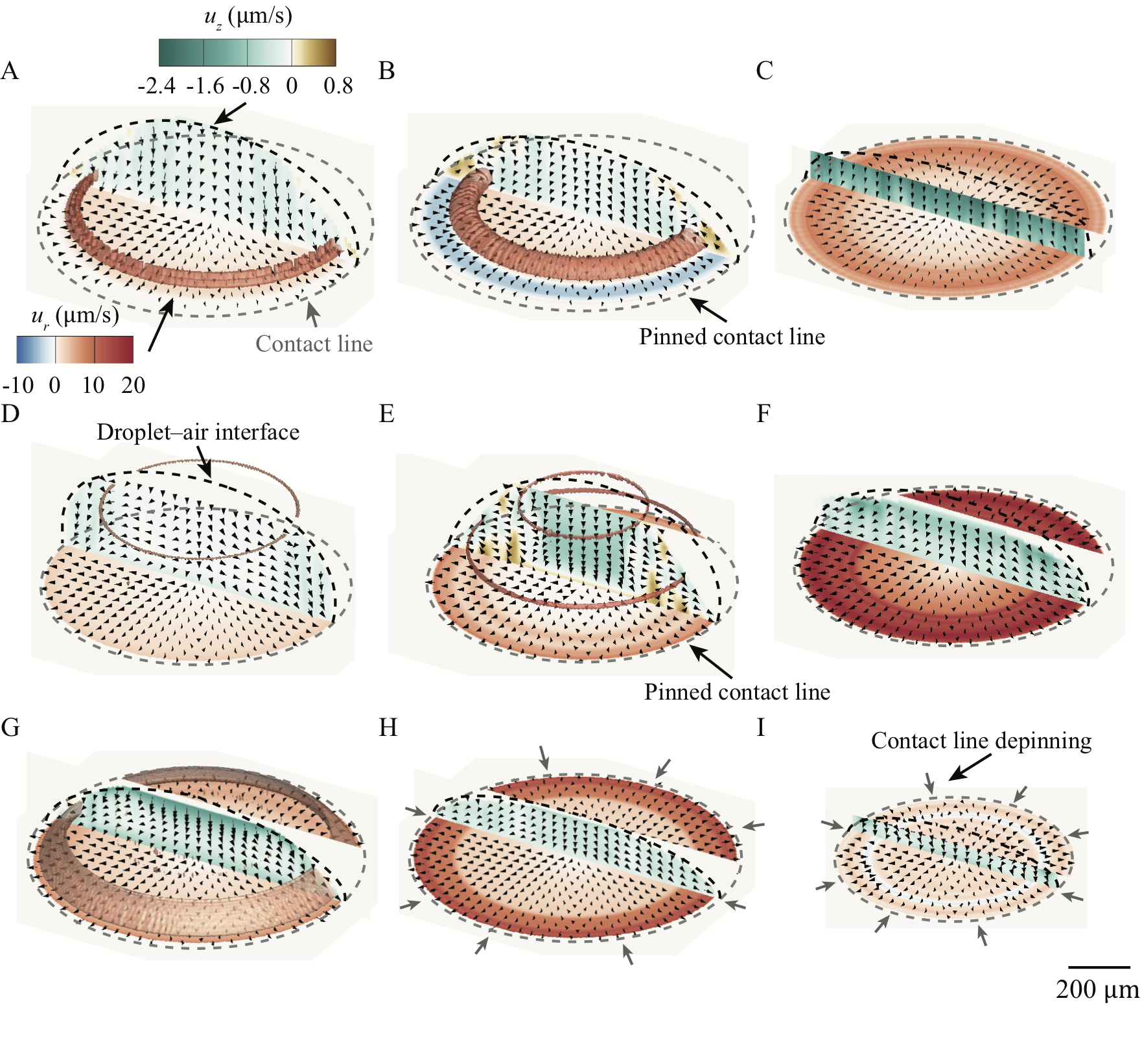}
\caption{The three-stage evolution of 3D internal flow for (A-C) a water droplet, (D-F) a sucrose solution (10 mM) droplet, and (G-I) an SDS solution (35 mM) droplet. The contour on the horizontal slice shows the radial velocity ($u_r$) measured near the substrate, and that on the vertical slice across the droplet center is the vertical velocity ($u_z$). The same legends apply to all the cases, with positive values indicating outward and upward flow and negative values representing inward and downward flow. The red iso-surface represents the vorticity magnitude $|\omega|=0.24 \mathrm{~s}^{-1}$ for visualizing Marangoni flows. The dashed lines represent the droplet boundary.}
\label{fig:4}
\end{figure}

During the evaporation process, the gradients of evaporation flux and surface tension (caused by gradients of surface temperature or chemical concentration) lead to the development of internal flows. Based on the relative dominance of capillary and Marangoni effects, the evolution of the internal flow is divided into different stages for the water droplet, i.e., Stage I with competing capillary flow and thermal Marangoni flow, Stage II with dominant thermal Marangoni flow, Stage III with a dominant capillary flow. Such three stages are further quantified using our experimental results and simulations showing the radial flow velocity variation throughout the evaporation process as shown in Fig. \ref{fig:7}. The 3D velocity field at Stage I (Fig. \ref{fig:4}A) shows the capillary flow dominates the majority portion of the droplet around the center. However, a recirculating Marangoni flow develops near the contact line (see the vorticity iso-surface in Fig. \ref{fig:4}A), i.e., inward flow near the droplet edge and droplet–air interface. As the internal flow evolves, at Stage II (Fig. \ref{fig:4}B), the Marangoni flow overtakes the capillary flow, covering a larger region extended from the contact line. A stronger and larger vortex ring associated with this Marangoni flow is illustrated by the vorticity iso-surface.  As contact line depinning begins (i.e., radial velocity decreases near the contact line in Fig. \ref{fig:4}C) at Stage III, the recirculation disappears due to the contact angle reaching the threshold of about 14\textdegree{} reported in \cite{hu2005b}, and the internal flow field is again dominated by the capillary flow like that at Stage I, which leads to the ‘coffee-ring’ deposition. At this stage, we observe an increased vertical velocity in the flow field across the droplet center (Fig. \ref{fig:4}C), leading to the downward migration of particles near the center, causing the inner deposition of particles. Throughout the entire evaporation process, our 3D flow measurements reveal the expansion of Marangoni flow inward, potentially due to increasing surface temperature gradient along the radial direction and the increase of vertical velocity, both of which contribute to the final particle deposition patterns.

For the sucrose solution droplet, the internal flow also undergoes a three-stage evolution (Fig. \ref{fig:4} D$\sim$F) and shows a different trend in comparison to that of water droplets. In particular, at Stage I (Fig. \ref{fig:4}D), the Marangoni flow reduces its strength with increasing sucrose concentration due to the surface tension gradient associated with changing chemical concentration (solutal Marangoni effect) in the opposite direction to that from the temperature gradient (thermal Marangoni effect). Accordingly, the weakening of the recirculating flow on top of the capillary flow causes a much narrower recirculation region than the water droplet (see the vorticity iso-surface in Fig. \ref{fig:4}D). Subsequently, at Stage II, the combined Marangoni flow expands inward near the droplet–air interface, similar to that in the water case but with a weaker recirculation, while the capillary flow becomes dominant near the substrate (Fig. \ref{fig:4}E). As the contact angle reaches the minimum, capillary flow dominates over the whole flow field at Stage III, and the vertical velocity is smaller than that in the water case (Fig. \ref{fig:4}F). Thus, such a reduced vertical migration relative to the radial migration towards the contact line of particles leads to a more prominent ‘coffee-ring’ deposition.

Finally, the internal flow of an SDS droplet also experiences a multi-stage evolution with trends different from those for water and sucrose droplets (Fig. \ref{fig:4} G$\sim$I). Specifically, we observe significantly stronger combined thermal and solutal Marangoni flow with increasing SDS concentration at Stage I of evaporation as, in this case, the chemically induced surface tension gradient is along the same direction as that due to the thermal effect. This enhanced Marangoni effect leads to a larger recirculation region within the droplet highlighted by the vorticity iso-surface (Fig. \ref{fig:4}G). The Marangoni vortex weakens at Stage II as the contact angle reaches the critical value ($14^\circ$ based on \cite{hu2005b}) earlier than those in the water and sucrose solution droplet, and the capillary flow is dominant over the whole flow field (Fig. \ref{fig:4}H). Such a dominant capillary flow carries particles to the receding contact line (earlier depinning), leading to scattered particle deposition along the way. As the droplet evaporates, at Stage III, we observe a suppressed outward capillary flow as it shows a much smaller and even inward radial flow in comparison with that at Stage II (see the horizontal slice near the substrate in Fig. \ref{fig:4}I). Such a suppression in capillary flow due to the moving contact line, as shown in \cite{masoud2009}, reduces the radial particle migrations towards the contact line, leading to the deposition of a higher concentration of particles around the droplet center.

Note that Fig. \ref{fig:4} only illustrates the most prominent features of the three-stage internal flow evolution at three time instances throughout the evaporation process for each droplet. The flow field may vary smoothly or abruptly between time instances, depending on the presence of unsteady variables or critical conditions. In Fig. \ref{fig:7}, we further quantify such flow field evolution using radial and vertical velocity evolutions at certain locations within the droplets. Specifically, the transitions between Stage I and Stage II of the water and sucrose solution droplets are associated with the development of temperature and concentration gradients along the droplet-air interface. Thus, at 21\textdegree{}C ambient temperature and 45$\%$ relative humidity, the time scales related to these two stages are typically around $0.3t_f \sim 0.4t_f$, where $t_f$ is the total evaporation time. The evolution from Stage II to Stage III for these two types of droplets is triggered by the critical contact angle ($14^\circ$ based on \cite{hu2005b}) of the recirculating Marangoni flow and contact line depinning. Thus, Stage III of water droplet takes around $0.2t_f \sim 0.3t_f$, while Stage III of sucrose solution droplet takes only $0.1t_f$ due to delayed contact line depinning. As for the SDS droplet, the rapid transition from Stage I to Stage II is also due to the critical contact angle and early depinning of the contact line that occurs at around $0.2t_f \sim 0.3t_f$. While the evolution from Stage II to Stage III in the SDS droplet is characterized by the interplay between capillary flow and flow induced by the inward-moving contact line, with the latter becoming dominant at around $0.85t_f$ in the current study.

Given the comprehensive 3D characterization of the internal flow motion, we further expand our analysis to include estimates of the characteristic dimensionless numbers. Specifically, we measure the Reynolds numbers ($R e=\rho \overline{u_r} R_0 / \mu$), the ratio between inertial forces and viscous forces, to be from 0.004 to 0.007, on the same scale as the value of 0.003 reported in \cite{hu2005a}. This small value is due to weak internal flow, suggesting that the inertial terms can be neglected in the N-S equation. The capillary number ($Ca=\mu \overline{u_r} / \sigma$) is the ratio between the viscous forces and surface tension force. Small capillary numbers, $Ca \sim O(10^{-7})$, suggest that the surface tension dominates normal stress balance, keeping the shape of the droplet. Moreover, the small Bond numbers, $Bo \sim 0.03$, suggest that gravitational forces can be neglected, and a spherical cap shape for the droplets can be maintained due to larger surface tension forces compared to the gravitational forces. Finally, we consider the combined
Marangoni number as the sum of the thermal and solutal Marangoni numbers: $M a=M a_T+M a_S=-\beta_T \Delta T_0 t_f / \mu R-\beta_S \Delta C_0 t_f / \mu R=\tau t_f / \mu$. We summarize the values of these dimensionless numbers in Table \ref{table1}. In these equations, $\overline{u_r}$ is the average radial flow velocity; $\mu$ is the viscosity; $\sigma$ is the surface tension; $\beta_T$ and $\beta_S$ are the temperature and concentration coefficient of surface tension; $\Delta T_0$ and $\Delta C_0$ are the surface temperature and concentration differences; $\tau$ is the average surface shear stress. Note that we do not directly measure the concentration and temperature gradient: $\Delta C_0$ and $\Delta T_0$. Instead, we calculate the shear stress at the droplet–air interface to represent the surface tension gradient \cite{marin2016surfactant}. These combined Marangoni numbers are also used as inputs for our predictive model.

\begin{table}
$$
\begin{array}{|c|c|c|c|}
\hline \text { Droplet system } & \text { Water } & \text { Sucrose } & \text { SDS } \\
\hline R e & 0.004 & 0.007 & 0.005 \\
\hline C a & 6.3 \times 10^{-8} & 9.5 \times 10^{-8} & 15.8 \times 10^{-8} \\
\hline M a & 50 & 40 & 170 \\
\hline B o & 0.029 & 0.034 & 0.033 \\
\hline
\end{array}
$$
\caption{Summary of the dimensionless numbers of the three droplet systems.}
\label{table1}
\end{table}

\subsection{\label{sec:33}Particle migration near the contact line}

\begin{figure}
\centering
\includegraphics[width=16cm]{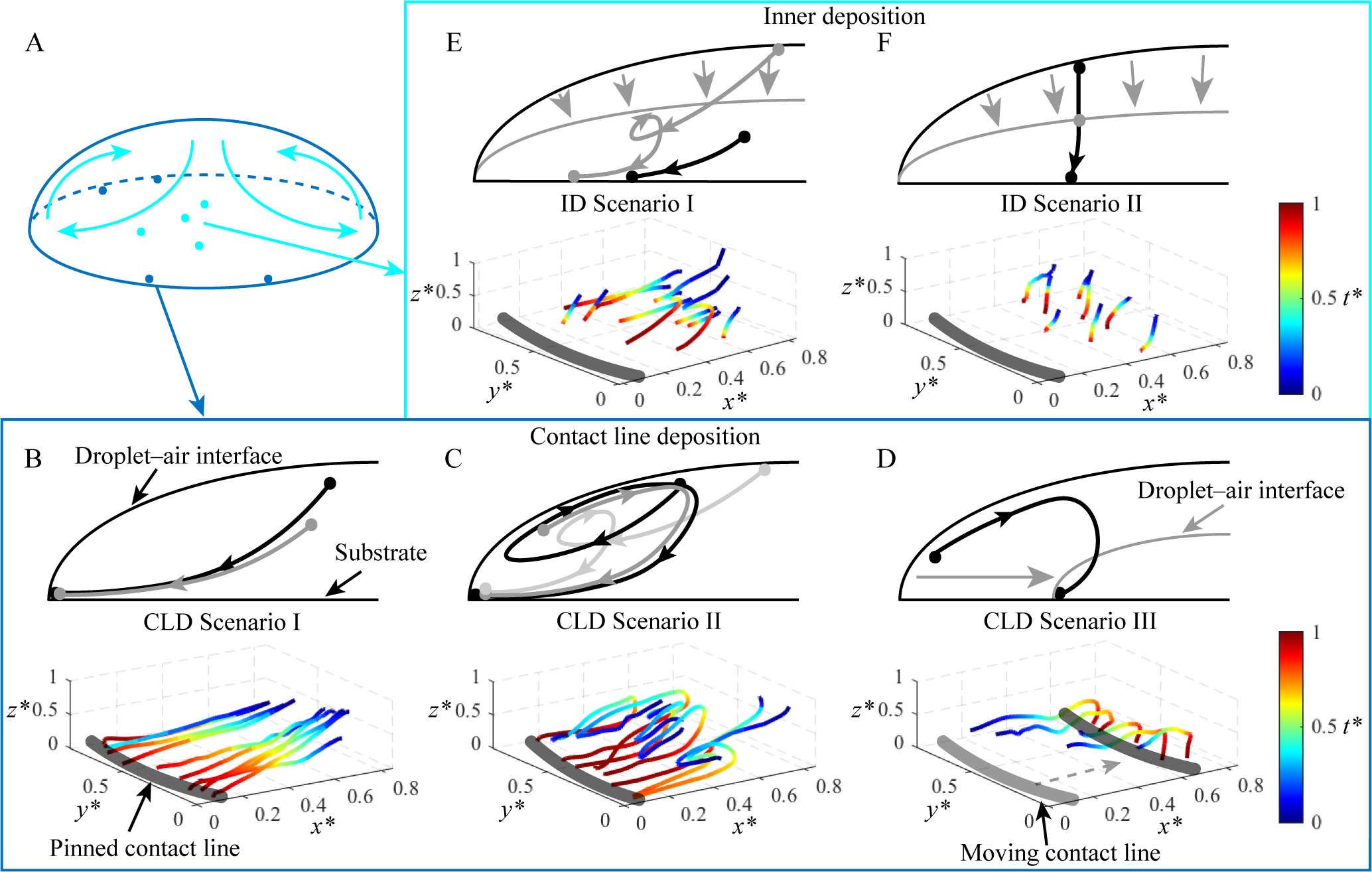}
\caption{Particle deposition scenarios. (A) A schematic showing the internal flow and contact line deposition (CLD, black dots) and inner deposition (ID, gray dots) in an evaporating droplet. The 2D illustration and the corresponding samples of particle trajectories for (B, C, D) different CLD scenarios and (E, F) ID scenarios. Different shades of colors (black and gray) represent different scenarios. Note that only a small subset of ~5–10 particle trajectories are selected for the clarity of the presentation.}
\label{fig:5}
\end{figure}

Using a higher magnification DIH imaging near the contact line, we are able to track the particle movement in this region during the entire evaporation process. Such measurements provide us with critical information to link the evolution of internal flow and the final deposition patterns. Based on our observation, we identify three scenarios that lead to contact line deposition (CLD) and two associated with inner deposition (ID) (Fig. \ref{fig:5}).

In the first scenario of CLD (i.e., CLD Scenario I, Fig. \ref{fig:5}B), the particles follow a smooth path from the center to the contact line. Such particle movements are strongly driven by the outward capillary flow, which has been regarded as the main cause of “coffee-ring” deposition. It occurs for particles located initially outside the influence zone of the Marangoni flow during evaporation. This scenario is observed in all cases but more often in sucrose cases where the capillary flow is the most dominant. In CLD Scenario II, the particles follow trajectories that exhibit reverse radial migrations at different stages of evaporation (Fig. \ref{fig:5}C), highlighting the highly unsteady internal flow during the evaporation. Specifically, the Marangoni flow alters the particle migration direction, while the prevailing capillary flow during the final stage redirects particles back toward the contact line. This observation is consistent with comparable findings by Marin et al. \cite{marin2019solutal}. The particle trajectories categorized in this scenario are present both in water and sucrose cases due to the rising and falling trend of Marangoni flow as the droplet evaporates and more often appear in the water case due to stronger Marangoni flow. Finally, in CLD Scenario III, the particles migrate inward near the droplet–air interface from a location close to the initial contact line and finally deposit downward and outward at the instantaneous contact line (Fig. \ref{fig:5}D). These particles are initially affected by the Marangoni flow and receding contact line migrating inward and eventually reversing direction due to the dominant capillary flow as the Marangoni flow diminishes. Particle trajectories in this scenario are found in the SDS solution droplets as such flow evolution, and early inception of contact line depinning is unique among the three types of droplets. 

For inner deposition, particles can follow similar trajectories as in CLD Scenario I and II but deposit inside the pinned contact line (ID Scenario I, Fig. \ref{fig:5}E). In comparison to CLD Scenarios, the particles are either initially located close to the substrate or experience more extensive downward migration in response to the more dominant vertical flow relative to the lateral flow. Accordingly, ID Scenario I is more often found in water cases though it occurs in all cases. In the other ID scenario (ID Scenario II), particles are trapped by the droplet–air interface due to the local equilibrium of interfacial forces, drag forces, etc. \cite{jung2010,joshi2010} and migrate vertically with the droplet–air interface with minimum radial movements (Fig. \ref{fig:5}F). These particle trajectories are more often observed in the water and sucrose solution droplets due to stronger surface tension forces to trap the particles than in the SDS solution droplet.

\subsection{\label{sec:34}Predictive model of droplet evaporation and particle deposition}

Based on the aforementioned experiment results, we demonstrate that the evolution of the internal flow and deposition pattern is influenced by the combined effects of capillary flow, Marangoni flow, and boundary movement. Therefore, we propose a predictive model for the deposition pattern based on previous analytical studies by \cite{hu2005a,hu2005b,masoud2009}, incorporating the combined effects into the simulation of internal flow and particle migration during the droplet evaporation processes. To enable analytical investigation, we model the droplets with uniform boundary conditions in the azimuthal direction. In addition, a 3D spherical cap shape for the droplet is assumed, considering the small Bond number. This simplification enables us to assume axisymmetric internal flow independent of the circumferential coordinate $\theta$ while allowing for particle motion in the circumferential direction, differentiating it from 2D models. It is a more accurate representation of the flow field and will lead to a better understanding of the underlying physics. 

With the input droplet parameters (i.e., change of droplet contact radius $R(t)$ and contact angle $\phi(t)$ with time) measured from the experiments, the model first simulates the internal flow (radial velocity $u_r$ and vertical velocity $u_z$) based on analytical solutions from continuity, Stokes, and energy (Laplace) equations derived from \cite{hu2005a,hu2005b} for capillary flow and Marangoni flow, and \cite{masoud2009} for flow induced by boundary movement. Specifically, 
for a droplet with contact radius $R$, contact angle $\phi$, droplet top surface height $h(r,t) = \sqrt{\left(\frac{R}{\sin \phi}\right) ^ 2 - r ^ 2 } - \frac{R}{\tan \phi}$, initial droplet height $h_0=h(0,0)=\frac{R_0}{\sin \phi} - \frac{R_0}{\tan \phi}$, and total evaporation time $t_f$, we define the internal flow driven by the evaporative flux as $(u_{r,CA}, u_{z,CA})$, and it is modeled from the beginning to the end of the evaporation process by the equations from \cite{hu2005a}:
\begin{equation}
\tilde{u}_{r,CA}=\frac{3}{8} \frac{1}{1-\tilde{t}} \frac{1}{\tilde{r}}\left[\left(1-\tilde{r}^{2}\right)-\left(1-\tilde{r}^{2}\right)^{\frac{\phi}{\pi}-\frac{1}{2}}\right]\left(\frac{\tilde{z}^{2}}{\tilde{h}^{2}}-2 \frac{\tilde{z}}{\tilde{h}}\right)
\end{equation}

\begin{equation}
\begin{multlined}
\tilde{u}_{z,CA}= \frac{3}{4} \frac{1}{1-\tilde{t}}\left[1+\left(\frac{1}{2}-\frac{\phi}{\pi}\right)\left(1-\tilde{r}^{2}\right)^{\frac{\phi}{\pi}-\frac{3}{2}}\right]\left(\frac{\tilde{z}^{3}}{3 \tilde{h}^{2}}-\frac{\tilde{z}^{2}}{\tilde{h}}\right)+ \\
\frac{3}{2} \frac{1}{1-\tilde{t}}\left[\left(1-\tilde{r}^{2}\right)-\left(1-\tilde{r}^{2}\right)^{\frac{\phi}{\pi}-\frac{1}{2}}\right]\left(\frac{\tilde{z}^{2}}{2 \tilde{h}^{2}}-\frac{\tilde{z}^{3}}{3 \tilde{h}^{3}}\right) \tilde{h}(0, \tilde{t})
\end{multlined}
\end{equation}

In these equations, $u_r$ and $u_z$ are normalized by the characteristic velocity $U=R⁄ t_f$, and other dimensionless parameters are defined as $\tilde{t}=t/ t_f$, $\tilde{r}=r/ R$, $\tilde{z}=z/ h_0$, and $\tilde{h}=h/ h_0$. In general, the capillary flow always points outward, and its speed increases with $r$ and $t$ for a pinned contact line. In our current model, we omit the higher-order terms within the braces of the original equations based on the lubrication approximation and the assumption of wetting droplets. The model could be extended to include non-wetting droplets with contact angles over 90 degrees by integrating additional equations for their internal flows, thus broadening its applicability to a wider range of droplet behaviors.

Based on the theory from \cite{hu2005b}, the internal flow caused purely by the Marangoni effect $(u_{r,MA}, u_{z,MA)}$ can be expressed as:
\begin{equation}
\tilde{u}_{r,MA}=\frac{M a h_{0} \tilde{h}}{2 R}\left(a b \tilde{r}^{b-1}+2(1-a) \tilde{r}\right)\left(\frac{\tilde{z}}{\tilde{h}}-\frac{3}{2} \frac{\tilde{z}^{2}}{\tilde{h}^{2}}\right)
\end{equation}

\begin{equation}
\begin{multlined}
\tilde{u}_{z,MA}=-\frac{M a h_{0}}{4 R}\left(a b^{2} \tilde{r}^{b-2}+4(1-a)\tilde{r}^2\right)\left(\tilde{z}^{2}-\frac{\tilde{z}^{3}}{\tilde{h}}\right)+ \\
\frac{M a h_{0}}{2 R}\left(a b \tilde{r}^{b}+2(1-a) \tilde{r}^{2}\right)\left(\frac{\tilde{z}^{3}}{\tilde{h}^{2}}\right) \tilde{h}(0, \tilde{t})
\end{multlined}
\end{equation}

The time-varying Marangoni number $Ma$ is used here, considering the integrated effects of temperature and chemical composition concentration gradients. The coefficients $a$, $b$, and $c$ are used to depict these gradients, based on values from \cite{hu2005b} and the contact angle from our experimental measurements. We model the velocities with $R(t)$ and $\phi(t)$ measured from the experiments, and the Marangoni number and coefficients are functions of the contact angle $\phi$ obtained from \cite{hu2005b}. The Marangoni effect causes recirculating flow within the evaporating droplet. Its strength decreases with the decreasing $Ma$ and $\phi$. The recirculating flow diminishes at the critical contact angle $\phi=14^\circ$.

We finally model the internal flow change caused by the boundary movement based on the geometrical change of the droplet following \cite{masoud2009}:
\begin{equation}
\tilde{u}_{r,CL}=\frac{\tilde{r}}{R(\tilde{t})} \frac{d R}{d \tilde{t}}
\end{equation}

\begin{equation}
\tilde{u}_{z,CL}= \frac{R\tilde{z}}{h_0}\frac{d \tilde{R}}{d \tilde{t}} \frac{1-\sin \phi}{\cos \phi} + \frac{R\tilde{z}}{h_0\phi} \left(\frac{\sin \phi - 1}{\cos ^{2} \phi} \right)\frac{d \phi}{d \tilde{t}}
\end{equation}

The radial velocity $u_{r,CL}$ is caused by the receding contact line at a rate of $\dot{R}$, and the vertical velocity $u_{z,CL}$ can be derived from the rate of change of the droplet height $\dot{h}$. Note that the rate of change of the contact angle $\frac{d \phi}{d \tilde{t}}$ is considered to be zero during the contact line receding motion. As the rate of change of the droplet boundary increases over time, the resulting internal flow usually accelerates.

With the assumption that the boundary conditions considering the three effects are linearly integrated, the corresponding analytical solutions are then combined in an additive manner $\boldsymbol{u}(r, z)=\boldsymbol{u}_{\mathrm{CA}}+\boldsymbol{u}_{\mathrm{MA}}+\boldsymbol{u}_{\mathrm{CL}}$. The internal flow within the droplet can be influenced by the interplay between these three different effects, leading to different flow directions at certain points within the droplet. Thus, the competition between these effects can be highlighted by examining the sign of the flow speed at these points. 

\begin{figure}
\centering
\includegraphics[width=10cm]{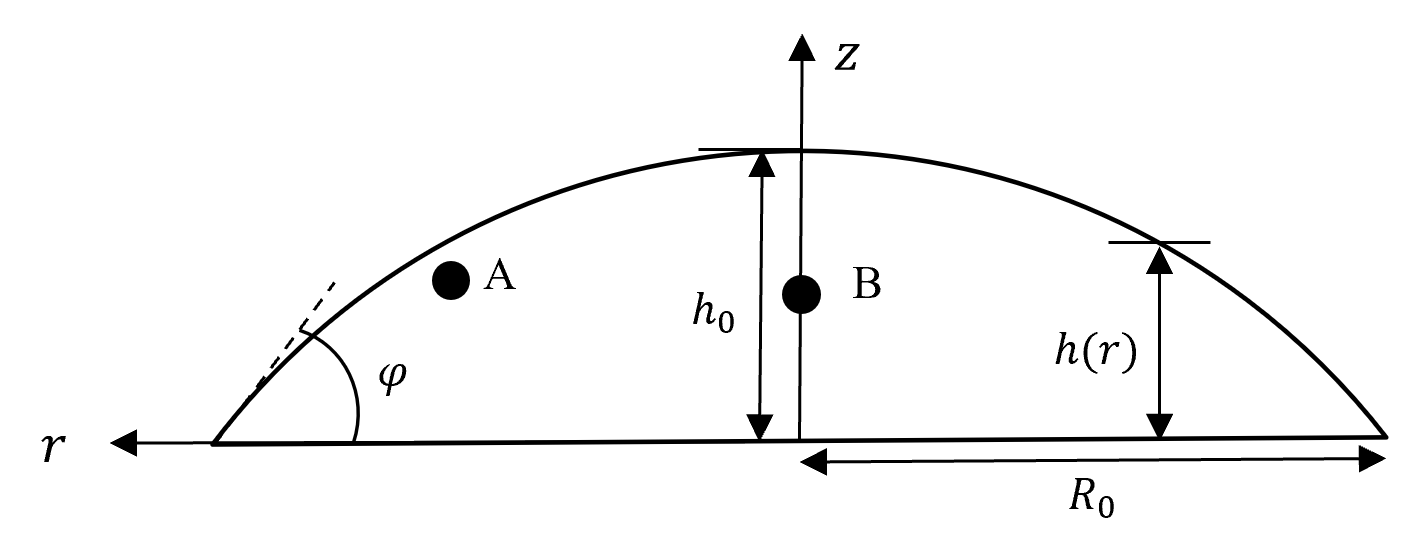}
\caption{A schematic of the simulation domain for the analytical internal flow model. Point $A$ is selected at $\tilde{r}=0.6$ and $\tilde{z}/\tilde{h}=0.85$ to represent the radial velocity evolution, and point $B$ is selected at  $\tilde{r}=0$ and $\tilde{z}/\tilde{h}=0.5$ to represent the vertical velocity evolution, as illustrated in the figure by the solid black circles.}
\label{fig:6}
\end{figure}

We show that our analytical model of internal flows is able to predict the evolution of internal flow correctly for the water, sucrose solution, and SDS solution droplets through a comparison of our experiment results (Fig. \ref{fig:7}A and B). We select two points $A$ and $B$ to illustrate such evolution, with point $A$ near the droplet–air interface for the radial velocity ($u_r$) and point $B$ at the center of the droplet for the vertical velocity ($u_z$). The radial velocity for the water droplet demonstrates the initial development of the recirculating flow, shown as $u_r$ becomes negative). Such a negative velocity is maintained for a longer period compared to the sucrose and SDS cases (Fig. \ref{fig:7}). The model also predicts a weaker Marangoni flow for the sucrose solution droplet that quickly evolves to the capillary flow dominant regime. Moreover, the inward flow due to contact line depinning for the SDS case is also captured by our model. {To validate the predicted change in the internal flow regime, a comparison with the experiments is conducted. We calculated the bin-averaged radial and vertical velocities of particles around the selected points using equations:
\begin{equation}
\bar{u}_{r, \text { meas }}=\left.\frac{1}{N} \sum_{i=1}^N u_{p, r, i}\right|_{\tilde{t} \pm \Delta \tilde{t}, \tilde{r} \pm \Delta \tilde{r}, \tilde{z} \pm \Delta \tilde{z}},
\end{equation}

\begin{equation}
\bar{u}_{z, \text { meas }}=\left.\frac{1}{N} \sum_{i=1}^N u_{p, z, i}\right|_{\tilde{t} \pm \Delta \tilde{t}, \tilde{r} \pm \Delta \tilde{r}, \tilde{z} \pm \Delta \tilde{z}}.
\end{equation}
where $u_{p,r,i}$ and $u_{p,z,i}$ are the velocity components for particles within the averaging bin with 20 frames time range $\Delta t$, $\Delta \tilde{r}=0.05$, and $\Delta \tilde{z}=0.1$. We normalize both model predictions and experimental measurements using the maximum values to eliminate the complexities from the experiments, such as certain asymmetry of the droplet shape and variations in the environmental conditions. The experimental measurements generally agree with the theoretical predictions showing the transitions of the flow regimes (Fig. \ref{fig:7} A and B). Large deviations of the experiments from the theoretical predictions for all cases are observed during the period $\tilde{t}=0-0.4$, which can be attributed to the weaker Marangoni effect in real situations by surface contamination, as suggested by \cite{hu2006}. The uncertainties from the inhomogeneity of the experiment conditions and tracer location measurement (especially the $z$ location measurement) would also contribute to the discrepancies.

\begin{figure}
\centering
\includegraphics[width=16cm]{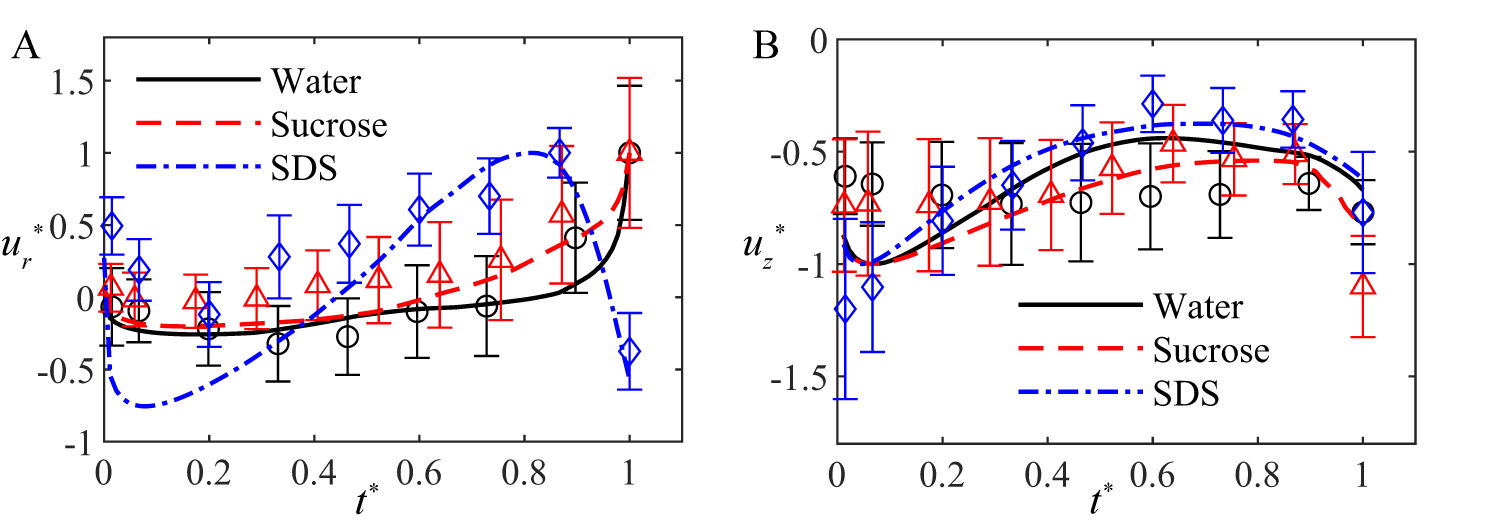}
\caption{(A) The comparison between the experiment measurements (symbols) and model predictions (lines) of the normalized radial velocity at the location $\tilde{r}=0.6$ and $\tilde{z}/\tilde{h}=0.85$ and (B) the normalized vertical velocity at the location $\tilde{r}=0.6$ and $\tilde{z}/\tilde{h}=0.85$ within the droplet at different time instants during the entire droplet evaporation. The experimental uncertainties (error bars) are quantified as the standard deviation of the scattered Lagrangian particle velocities within the region around the selected points.}
\label{fig:7}
\end{figure}

Subsequently, particles are randomly seeded in the simulated 3D droplets to follow the internal flow and Brownian motion. Such a particle migration model is based on the model proposed by \cite{kolegov2019}. In their model, the particle motion is confined in the simplified 2D droplet (only the $(r,\theta)$ plane). Thus, only an average radial velocity by capillary effect is considered, together with the particle attraction by capillary forces and diffusion due to Brownian motion. Our proposed method differs from that by including a more detailed consideration of the 3D internal flow caused by the three effects (i.e., $\boldsymbol{u}(r, z)=\boldsymbol{u}_{\mathrm{CA}}+\boldsymbol{u}_{\mathrm{MA}}+\boldsymbol{u}_{\mathrm{CL}}$), rather than using an average radial migration velocity. The corresponding particle advection within the time step $\delta t$ is $\delta l_{flow}=\left( \delta r, \delta z \right)=\left(u_r,u_z\right) \delta t$. Note that although the internal flow field is independent of the azimuthal direction $\theta$, we simulate the diffusion (i.e., Brownian motion) considering any direction in 3D defined by the random angles $\left(\alpha,\beta\right)$ in the range of $\left(-\pi,\pi\right]$. Thus, the particle movement by diffusion is $\delta l_{diff}=\sqrt{2D\delta t}$, where $D=\frac{k_{B} T}{3\pi \eta d}$ is the diffusion coefficient of the colloidal particles in the droplet, $k_B$ is the Boltzmann constant, $T$ is the absolute temperature, $\eta$ is the dynamic viscosity, and $d$ is the diameter of the particles. The subsequent directional displacement is defined as $(\delta x, \delta y, \delta z)=\left(\delta l_{\text {diff }} \cos \alpha \cos \beta, \delta l_{\text {diff }} \cos \alpha \sin \beta, \delta l_{\text {diff }} \sin \alpha\right)$. The simulated particles experience both advection by internal flow and diffusive motions at each time step yet are restricted by the contact line and the droplet-air interface. If the particle radial location $r_p$ is larger than $R_{lim}$ after flow advection and diffusion, where $R_{lim}$ is the radial location where the droplet height is the same as the particle diameter $d_p$, the contact line deposition happens. If the vertical location of a particle is smaller than half the particle diameter ($z_p \leq d_p/2$) within the contact radius ($r_p < R_{lim}$), we consider it to be inner deposition. Finally, we predict the particle deposition pattern as the number concentration of deposited particles along the radial direction $C_{\text {depo }}(r)$.

\begin{figure}
\centering
\includegraphics[width=15.5cm]{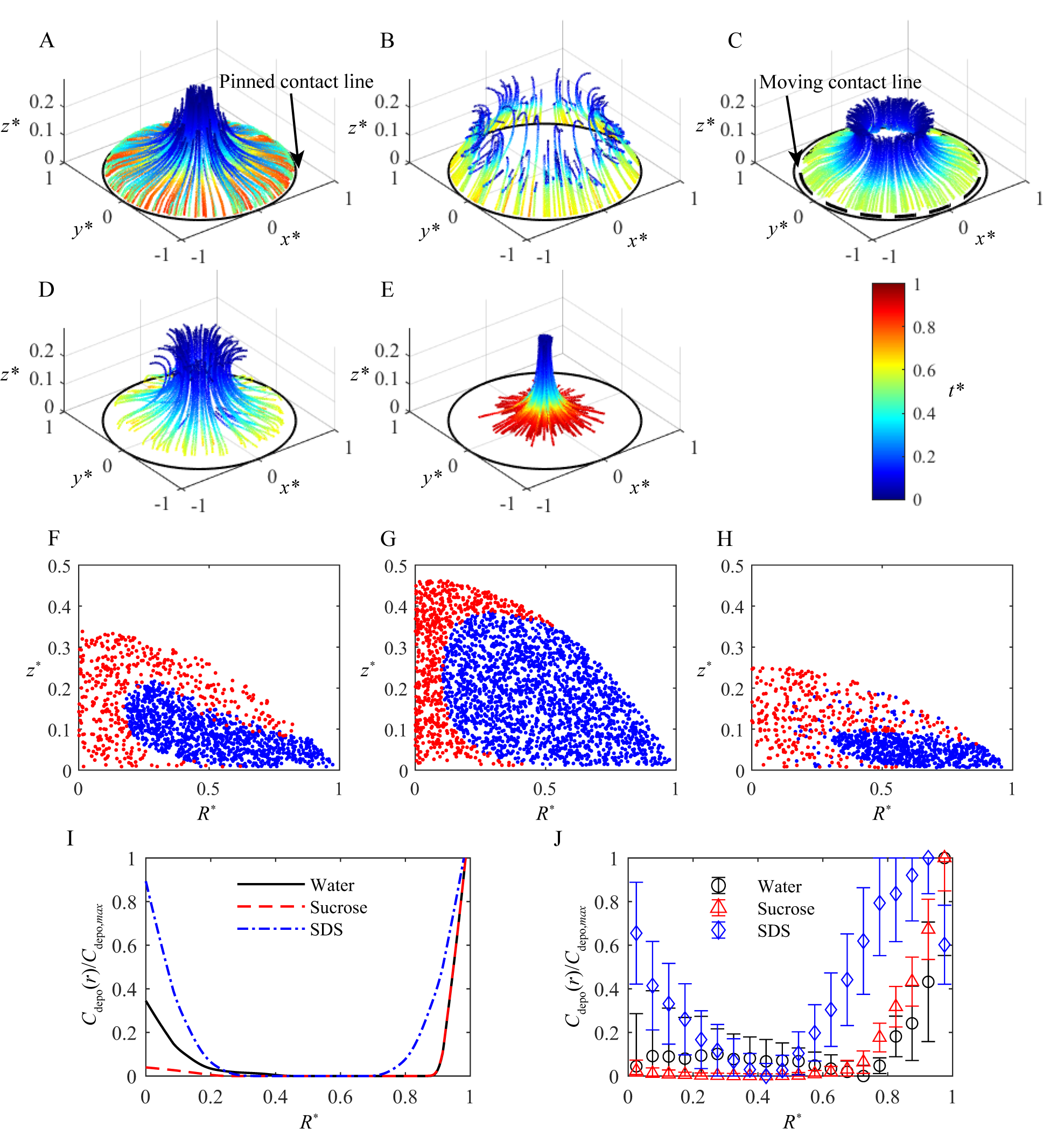}
\caption{Particle migration trajectories and deposition patterns derived from the predictive model compared to the experiments. The migration trajectories of seeded particles from their initial location to deposition simulated by our model for (A) The CLD Scenario I, (B) CLD Scenario II, (C) CLD Scenario III, (D) ID Scenario I, and (E) ID Scenario II. The initial particle distribution leads to contact line deposition (blue dots) or inner deposition (red dots) for (F) a water droplet, (G) a sucrose solution droplet, and (H) an SDS solution droplet. The normalized particle deposition concentration along the radial direction (deposition profile) from the (I) model prediction and (J) experimental measurements for the water droplet (blue line and circles), a sucrose solution droplet (red dashed line and triangles), and an SDS solution droplet (magenta dash-dotted line and diamonds). The deposition profiles are predicted as the number concentration of particles along the radial direction by the proposed analytical model, and measured using the image grayscale intensity of the phase projections of the holograms.}
\label{fig:8}
\end{figure}

Our particle migration model is then able to capture the five scenarios of particle migration trajectories summarized in Fig. \ref{fig:5} that lead to different particle deposition considering the specific evolving internal flow and droplet contact line motions. In addition to these mechanisms, the particles' the initial positions within the simulated droplets also significantly affect their final deposition locations, whether at the periphery or interior (Fig. \ref{fig:8} A$\sim$E). Specifically, the particles initially within $0.2 R_0 \sim 0.3 R_0$ annulus region would stay out of the recirculatory flow region, following capillary flow to deposit at the contact line (Fig. \ref{fig:8}A, corresponding to CLD Scenario I). Particles closer to the contact line ($0.6 R_0 \sim 0.8 R_0$ annulus region) and droplet–air interface experience the inward Marangoni flow and show reverse migration due to the internal flow evolution, leading to contact line deposition eventually by dominant capillary flow (Fig. \ref{fig:8}B, corresponding to CLD Scenario II). For the SDS solution droplet, particles within the $0.4 R_0 \sim 0.5 R_0$ annulus would follow the trajectories categorized in CLD Scenario III (Fig. \ref{fig:8}C) with the moving contact line. Compared to the predicted particle trajectories of CLD scenarios, the trajectories of ID scenarios show a larger slope in $z$. Specifically, particles within $0.1 R_0 \sim 0.2 R_0$ annulus region could experience larger vertical migration leading to inner deposition (Fig. \ref{fig:8}D, corresponding to ID Scenario I). In addition, particles initially located within 10\% of the droplet thickness from the substrate also have a much higher chance to deposit inside the contact line (Fig. \ref{fig:8}D). Although interfacial forces are not considered in our predictive model, particles initially located in close proximity of the droplet center (within $0.08 R_0$) and near the top interface (5\% of the droplet height) would experience minimum internal flow and migrate down with the droplet–air interface (Fig. \ref{fig:8}E, corresponding to the ID Scenario III). Based on the above analysis, particles initially located in different regions can follow different pathways to either CLD or ID. 

Furthermore, by grouping particles according to their deposition locations in the droplet (CLD or ID), we find clear demarcation (i.e., initially closer to the contact line or near the center, Fig. \ref{fig:8}F and G) of particle initial locations for the water and sucrose solution droplets with pinned contact line. We also notice the demarcation boundary between CLD and ID initial locations is a little fuzzier (i.e., some red dots appear amid blue dots) for water droplets than that for sucrose solution, potentially due to the early inception contact line depinning. While for the SDS solution droplet, due to the constantly moving contact line, the demarcation line on the initial distribution of CLD and ID particles is less clear, and there is a transition region between them where the deposition location of particles is sensitive to their initial positions (Fig. \ref{fig:8}H).

Moreover, with the particle trajectories throughout the evaporation process predicted, we can obtain the final deposition pattern $C_{\text {depo }}(r)$, which is represented using the number concentration of deposited particles along the radial direction (Fig. \ref{fig:8}I). Specifically, for water droplets, our model correctly predicts the mixed deposition pattern (i.e., “coffee-ring” deposition with inner deposition). As for sucrose, the accurately predicted particle deposition exhibits the most prominent “coffee-ring” pattern with much less inner deposition as compared to the water case. In addition, the model predicts a thicker band of deposition from the initial contact line and the most prominent inner deposition for the SDS case. For comparison, we extract the experimental deposition profiles (Fig. \ref{fig:8}J) from the phase projections (Fig. \ref{fig:3}D$\sim$F) of the deposition holograms \cite{minetti2008fast,yuan2008mapping,raffel2018particle}. These phase projections are derived by first capturing the phase information of the holograms at various reconstruction planes along the $z$ direction (perpendicular to the imaging plane). This collection of phase data is then projected onto a single plane using the maximum phase value, translating the $0 \sim 2 \pi$ scale into a grayscale image. Due to the difficulties in accurately separating the deposited particles, we employ annulus bins at varying radial positions and use average image grayscale intensity within each bin to determine the mean particle concentration. The experimental measurements show thicker rings for the water and sucrose solution droplets mainly due to the non-spherical droplet geometry. The thicker band deposition of the SDS solution droplet in the experiments is caused by the crystallization deposition overshadowing the particle deposition. In addition, the inner deposition of water droplets from experiments exhibits large uncertainty caused by mixed evaporation modes near the end of the evaporation process. Such an effect was smoothed in the predictive model, leading to underprediction of the inner deposition. 

Finally, we demonstrate the generalizability of our predictive model for other Newtonian droplets using a selected case of an isopropanol droplet where Marangoni flow dominates throughout the evaporation process. The temporal evolution of the droplet contact radius and contact angle is obtained from \cite{bhardwaj2009}. Such a dominant Marangoni flow is predicted by our model (Fig. \ref{fig:9}A), showing decreasing average speed as demonstrated by other simulations in \cite{hu2006} and \cite{bhardwaj2009}. We then simulate particle motions driven by the evolving internal flow and Brownian motion with a randomly distributed initial particle field within the droplet, the same as for the experimented droplets. Our model is then able to predict the final deposition pattern (Fig. \ref{fig:9}B), showing particles mostly deposited at the center of the droplet. Such a center deposition has been reported in previous simulations and experiments by \cite{hu2006} and \cite{bhardwaj2009}.

\begin{figure}
\centering
\includegraphics{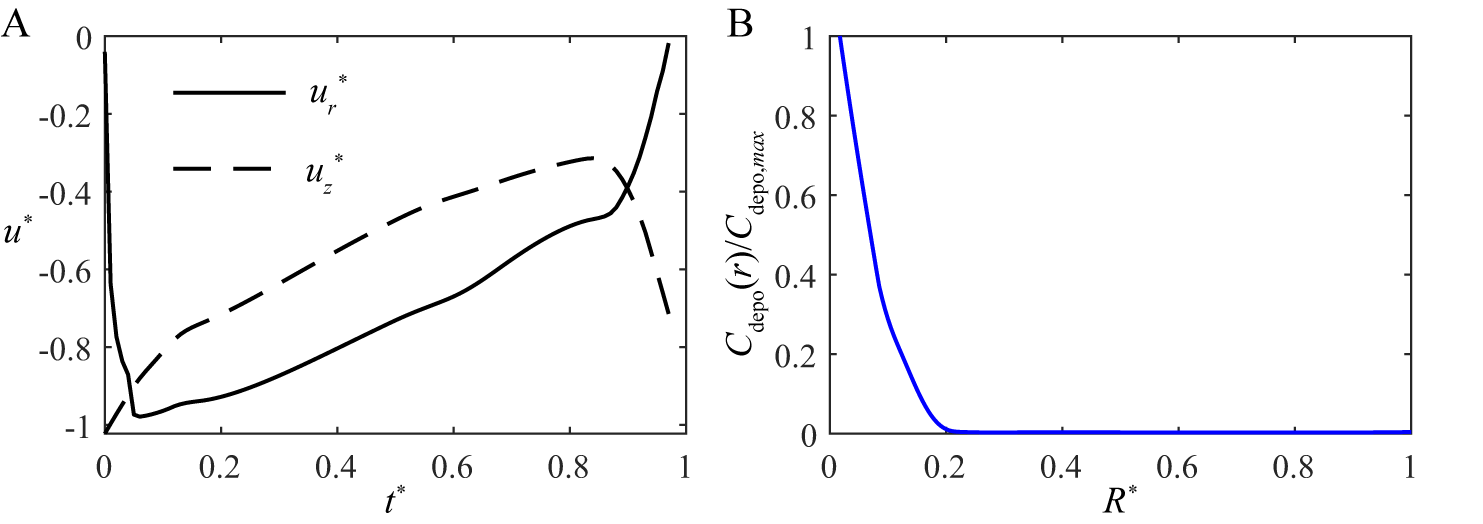}
\caption{Internal flow evolution and final deposition pattern of isopropanol droplet generated from our predictive model. (A) The predicted evolution of the normalized radial velocity and vertical velocity of the internal flow at the location $\tilde{r}=0.6$ and $\tilde{z} / \tilde{h}=0.85 \left(u_r\right)$ and location $\tilde{r}=0$ and $\tilde{z} / \tilde{h}=0.5 \left(u_z\right)$ within an isopropanol droplet and (B) the final deposition pattern as the normalized particle number concentration along the radial direction.}
\label{fig:9}
\end{figure}

\section{\label{sec:4}Conclusions}
In conclusion, our study provides a comprehensive analysis of the evolution of internal flow fields and particle migration for different types of droplets (water, sucrose solution, and sodium dodecylsulfate solution) throughout the entire evaporation process using a 3D imaging technique based on digital inline holography and particle tracking velocimetry (DIH-PTV). We demonstrate the three-stage evolution of the 3D internal flow regimes associated with the changing dominance of capillary flow and Marangoni flow, as well as droplet boundary movement during evaporation. Specifically, our results show the evolution of Marangoni flow both laterally and vertically within the evaporating droplets and the reduced outward flow and large downward fluid motions due to the contact line motion. Furthermore, we observe the changing migration directions of particles due to competing Marangoni and capillary flows during droplet evaporation. The analytical model complements our experimental findings by predicting the internal flow evolution of the droplets and deposition patterns with input droplet and particle properties. Such a model also determines the dependence of the deposition mechanisms of particles on their initial locations and the evolving internal flow field. We show that the internal flow evolution is critical to particle deposition patterns, and the model predictions are consistent with the experimental measurements. 

Our findings highlight the importance of incorporating the temporal evolution of the flow field into understanding the final particle deposition for colloidal droplets. The proposed analytical model can be potentially used for real-time prediction of the deposition pattern for feedback loop control involved in various industries, including spray painting, inkjet printing of micro-scale optical/electronic devices, chemical/biological sensors, and micromechanical/microfluidic devices.

Compared to previous studies that mainly focus on 2D observations \cite{manukyan2013, kim2016, mandal2012evidence, park2016, amjad2017}, our DIH imaging approach provides a more comprehensive view of the flow field. It also provides more accurate measurements in the direction perpendicular to the imaging plane compared to other 3D measurement techniques \cite{rossi2019interfacial, straub2021}. The DIH technique is versatile for measuring even smaller colloidal droplets with larger magnification. It is capable of measuring colloidal particles down to $\sim O(10^2)$ nm, limited by the diffraction limit.

The analytical model can be expanded to incorporate complex evaporation modes (boundary movement) into simulations using the finite element method or Lattice Boltzmann method to better understand the flow-particle interaction in realistic cases. Specifically, as most evaporating colloidal droplets exhibit non-Newtonian properties \cite{lu2016}, the model can be further improved by incorporating the constitutive equation between the viscosity and shear stresses for the non-Newtonian colloidal droplets into the internal flow simulation for more accurate predictions. These improvements will allow us to consider the effects of the complex droplet properties, surface properties, and environmental factors on relevant variables such as contact line movement, surface tension gradient, and evaporation flux, which can impact both Marangoni flow and capillary flow.

\begin{acknowledgments}
The authors would like to thank Prof. Howard A. Stone for inspiring discussions and suggestions and thank Ms. Chinmayee Panigrahi for her help in the experiments during the early stage of this research.
\end{acknowledgments}

\setcounter{section}{0}
\renewcommand{\thesection}{S-\Roman{section}}
\setcounter{figure}{0}
\renewcommand{\thefigure}{S\arabic{figure}}

\section{Supplementary Information}
\subsection{\label{sec:S1}Droplet evaporation behavior}

We measure the contact radius and contact angle of the droplets from the top view using digital inline holography (DIH) imaging and the side view employing regular bright field imaging. A spherical cap shape of the droplet is assumed considering the small Bond number ($\sim 0.03$). The contact radius can be easily measured from both the top view and the side view. While the contact angle is measured as the angle between the tangent line of the droplet–air interface at the contact point and the horizontal substrate. In Fig. \ref{fig:S1}, we have shown the change of contact angle in time for the three droplet systems. Moreover, the evaporation curves in Fig. 3 in the manuscript, which depict the change in the contact area of the droplets, are derived from the average of eight repeated experiments for each droplet system. The error bars in the graph represent the range of fluctuation. As a result, the mixed CCR and CCA mode for the water droplet may not be immediately evident in the averaged evaporation curve. To provide a clearer picture, we present in Fig. \ref{fig:S1} a single experiment of water evaporation that distinctly showcases the mixed mode.

\begin{figure}
\centering
\includegraphics[width=12cm]{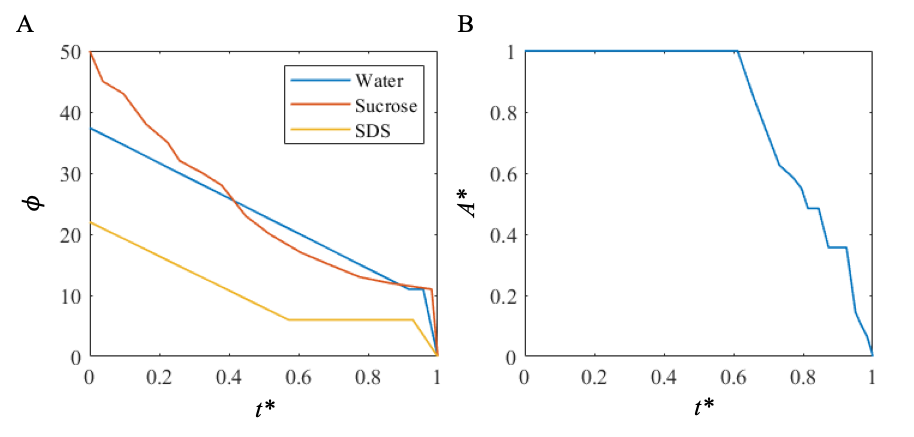}
\caption{(A) The variation of contact angle over time for the water, sucrose solution (10 mM), and SDS solution (35 mM) droplets. (B) Change of contact area over time for one water droplet case. The steps in the curve indicate the stick-slip mode of evaporation.}
\label{fig:S1}
\end{figure}

\subsection{\label{sec:S2}Marangoni number of the evaporating droplets}

We measure the Marangoni number of the combined thermal and solutal effects: $M a=M a_T+M a_S=-\beta_T \Delta T_0 t_f / \mu R-\beta_S \Delta C_0 t_f / \mu R=\tau t_f / \mu$. Instead of using the equation defined using $\beta_T$, $\beta_S$, $\Delta T_0$, and $\Delta C_0$, we calculate the shear stress at the droplet–air interface to represent the surface tension gradient for determining Marangoni number. Thus, we could not separate the thermal and solutal ones. Potentially, by assuming the same thermal properties among the three droplet systems and with the water droplet case as the baseline, we could extrapolate the thermal Marangoni number for the sucrose and SDS cases, and obtain the solutal Marangoni number by subtracting from the overall Marangoni number as the two are linearly integrated. For example, considering the based line thermal Marangoni number of 50, the solutal Marangoni number for sucrose is then $Ma_{S,suc}=-10$, while solutal Marangoni number for SDS is $Ma_{S,SDS}=120$.

However, the temperature profile and concentration profile is still used in the calculation of Marangoni flow in our model, especially the coefficients $a$, $b$, and $c$. Specifically, the temperature profile is defined as $T/ \Delta T = a \tilde{r}^b + \left( 1 - a \right) \tilde{r}^2 + c$. Through analogy, we can also assume a concentration profile: $C/ \Delta C = a^\prime \tilde{r}^{b^\prime} + \left( 1 - a^\prime \right) \tilde{r}^2 + c^\prime$, as both temperature and concentration distributions are governed by the diffusion equations. In our model, we assume a combined quantity $M$ that leads to the combined thermal and solutal Marangoni flows. The profile of this quantity: $M/ \Delta M = a \tilde{r}^b + \left( 1 - a \right) \tilde{r}^2 + c$ is used, and the coefficients $a$, $b$, and $c$ are determined based on the table from \cite{hu2005b} and the measured Marangoni number from the shear stress analysis. As both the concentration and temperature profiles evolve over time, the Marangoni number is also a variable that experience temporal variation. In Fig. \ref{fig:S2}, we demonstrate the temporal evolution of the velocity contributions from the capillary effect, combined thermal and solutal Marangoni effects, and contact line movement effect.

\begin{figure}
\centering
\includegraphics[width=16cm]{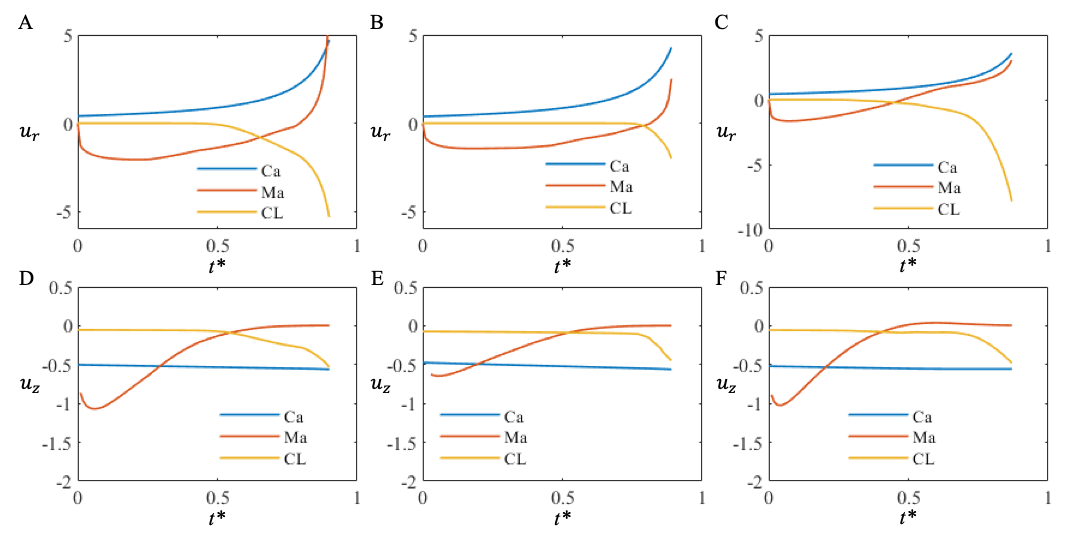}
\caption{The temporal evolution of the internal flow velocity at the selected points from Fig. 6. The velocities are normalized by the nominal velocity $U_0 = R_0⁄ t_f$. From left to right, the plots show the radial velocity evolution at point A for (A) water, (B) sucrose solution, and (C) SDS solution droplets, and (D, E, F) the vertical velocity evolution at point B.}
\label{fig:S2}
\end{figure}


\bibliography{PRF_droplet_evaporation_arxiv}

\end{document}